\documentclass[pre]{revtex4}
\usepackage{graphicx}
\usepackage{latexsym}
\usepackage{amsmath}
\usepackage{amssymb}
\usepackage{amsfonts}
\usepackage{amsthm}
\begin{document}
\title{ Controlling near shore nonlinear surging  waves through bottom
boundary conditions}
\author{ Abhik Mukherjee
\footnote{abhik.mukherjee@saha.ac.in}
, M.S. Janaki
\footnote{ms.janaki@saha.ac.in}
, Anjan Kundu
\footnote{anjan.kundu@saha.ac.in}
}
\affiliation{Saha Institute of Nuclear Physics\\
 Kolkata, INDIA}

\begin{abstract}
Instead of taking the usual passive view for warning of near shore surging
waves including extreme waves like tsunamis, we aim to study the possibility
of intervening and controlling nonlinear surface waves through the feedback boundary
effect at the bottom. It has been shown through analytic result that the
controlled leakage at the bottom may regulate the surface solitary wave
amplitude opposing the hazardous variable depth effect. The theoretical
results are applied to a  real coastal bathymetry in India.


\end{abstract}

\maketitle


\section{Introduction:}

Near shore coastal regions often witness surging of the approaching waves including
extreme events like tsunamis \cite{Kundu}. Such a natural phenomenon has also been
observed, though in a miniature scale in few rivers around the world as bore
waves \cite{1bore1,1bore2,1bore3}. Famous examples are the river Seine in France and the river Hoogli
 in India \cite{1bore1,Kundu}.
Such surging waves are suspected to be caused by nonlinear
 gravity waves, propagating over a decreasing depth bathymetry
towards the shore  or along upstream river. Such events which can often
trigger extremely hazardous effects have attracted intense attention over
centuries and have been studied extensively from both theoretical and
practical points of view \cite{Kundu}. The main emphasis of the investigations was to work
towards the development of early warning systems for minimizing  the loss
of human lives. The present development of the tsunami warning system has
definitely been reached to a satisfactory level\cite{1warning1}-\cite{1warning6}.

However, there are few situations where the installation of 
a passive warning system is not enough, while the demand is for more active
intervention. This is particularly true for  example, in protecting nuclear
reactors and related installations, which are located usually at the vicinity
of the sea shore due to logistic reasons, against the tsunami threat. As we
know the tsunami of 2004 which played  devastating effects spreading over
many countries was a potential threat to the nuclear reactor at Kalpakkam in
India. The tsunami of 2010  inflicted real calamities in Fukushima nuclear
reactors in Japan \cite{1Fukushima1,1Fukushima2}.

 In a relatively smaller scale, the near shore waves and
bore waves caused many devastating effects to the coastal habitats and
in-land rivers throughout the centuries . Therefore along with the traditional
warning systems, it is desirable to find ways and means geared towards
possible invasive procedures for taming of such hazardous wave phenomena.
There are few suggestions for effective interventions, like plantation of
Mangrove treas along the coastal lines \cite{2Mangrove1}, installation of breakwaters at
strategic positions \cite{2breakwater1}-\cite{2breakwater4}, stoppage of erosion by concrete bolders etc .

 However,
these are mostly indirect ways to counter the surging waves, while we lack
 proposals on directly attacking the problem, perhaps with the exception 
of the proposed bubble method, aiming to stop the incoming waves by a stream
of fast and strong counter-waves, mixed with air bubbles \cite{2Taylor}. Though the last
method was proposed more than fifty years back, its feasibility and
effectiveness has not been established yet.
The attenuation of incident water waves by a curved vane like 
structure positioned beneath or at the surface of a body of water is described in a Patent \cite{2control1}
where the detailed design of the structure is given. An attempt
was made to reduce the devastating effects
of a tsunami waves by single and double submerged barrier was done in Tel Aviv University\cite{2control2}.
They performed their experiments  in a basin 5 m in length and
10.5 cm in depth. The wavelength of the generated wave was
about 3 m, which allows referring to it as a tsunami. 

Our aim here is to put forward
an innovative proposal based on a theoretical study on the effect of a
feedback boundary control at the bottom on the surging surface wave
amplitude. The governing nonlinear equations describing unidirectional
gravitational waves are derived from the basic hydrodynamic 
equations at the shallow water regime. The key factor responsible for  surging of the 
 nonlinear waves approaching to the shore (or in upstream rivers ) is the
decreasing depth bathymetry, which triggers the amplitude surge of the
surface waves inversely proportional to the water depth which diminishes
continuously along the wave propagation towards the shore. 

Our strategy is to study first the effect of the bottom boundary condition
on the  nonlinear solitary  surface waves of the well-known perturbed KdV equation,
propagating in shallow water of constant depth. The vertical fluid velocity at the
bottom is taken as a function of  surface wave profile, to identify subsequently
through theoretical analysis,  the optimal case
inducing maximum amplitude damping to the surface waves.
This knowledge is applied through slowly varying bathymetry, which without
the leakage condition, as we know would result to solitary wave solution
with increasing amplitude with the water depth decreasing along its
propagation. However when the controlled bottom leakage with optimal
feedback wave profile is imposed, the surging amplitude of the wave meets
the counter damping effect, resulting to a managed propagating waves towards
the shore with reduced hazardous effect due to the effective damping of the
wave.

 We would like to emphasize that there could be various natural bottom
boundary effects inducing damping of the surface wave amplitudes, like porosity\cite{Murray}-\cite{boussinesq2},
irregularities , uneven heights, periodic topography, friction\cite{Ostrovsky,Pelinovsky} apart from the
fluid viscosity \cite{OttSudan2} etc. while the long obstacle can induce fission of the
solitary waves \cite{Ostrovsky}. However our aim here is to induce damping effect
artificially through controlled mechanism.

The privilege of our theoretical result is the exact nature of the
solutions we obtain, in spite of the variable depth bathymetry, which is
rather a rare achievement. Our theoretical results with exact solutions
allow to extract finer details and precise predictions.
Our findings are extended to cover different cases of the controlled bottom
leakage conditions, ranging from space dependent to time dependent, from
vanishing of effective leakage velocity to a desirable leakage conditions
etc. Our theoretical findings for the possible  control of the surging waves like
tsunamis and bore waves based on our exact results are applied next to
real sea shore bathymetries. We have focused in particular on two high risk
coastal zones of bay of Bengal near the city in Chennai of south India as
presented in a recent in depth study of the subject \cite{ArunKumar}.
Our analysis shows that a significant upsurge could have been experienced by
a future surging wave  approaching towards these coastal points. For
example at the identified northern coastal point (N $13^\circ$ $10.5'$ - E $80^\circ$ $18.75'$) a wave of nearly 1
meters built at a distance of 10.5 km from the shore would have been developed
to a killing height of 30 meter at the shore without any control. Similarly
at a southern point (N $13^\circ$ $0'$ - E $80^\circ$ $16.2'$)
 the bathymetry would induce equally devastating upsurge
for a wave of  1 m created at 11.1 km away, to develop into a 30 meter killer
wave at the shore. Applications of feedback controlled method
through bottom boundary condition, that we propose here is found to be able
to regulate such upsurging waves to a considerable extent, minimizing its
hazardous effects. In particular, the  surging waves at the northern point
could be regulated at the height of 1.23 meters if the leakage installation
could be made starting from a distance of 900 meters from the shore. A
smaller distance could result to a higher amplitude though significantly
lesser than that without control.
 Similarly the surging waves at the
southern coast could be controlled to a wave amplitude of 0.4 meter, if the
installation starts at a distance of 900 meters. Thus the hazardous effects
of tsunami like surging waves could possibly be neutralized to some extent
through a controlled bottom leakage condition tuned by a linearly dependent
wave profile which we found to be optimal, created through a feedback mechanism. 

 The paper is organized as   follows.  In section II, the shallow water
problem in constant depth with a specific class of feedback bottom BC with
controlled leakage is considered. Corresponding 
 surface wave evolution  equation is derived for different leakage
conditions and solved for its
 solitary wave solutions,  
 through Bogoliubov-Mitropolsky approximation.
 In the next section variable depth problem is taken up, for the optimally
controlled leakage and the related wave equation is derived extracting 
exact solution  for a tuned balance between the effects of leakage  and the
variable bathymetry.

Our theoretical findings are applied to a real near shore bathymetry data
and our predictions are checked for different installation distance for the
controlled mechanism which is contained in sec IV.
The extension of this exact result  is considered in the appendix ( sect. VI) for a particular bathymetry 
and leakage velocity obtaining again  a solvable equation
with exact solution. This section also includes
time dependent leakage and leakage that contains both wave profile dependent and 
independent part.
 Section V is the concluding section followed by the bibliography.


\section {Effect of feedback bottom boundary condition on nonlinear surface
wave in constant depth}
\begin{figure}[!h]
\centering

{
 \includegraphics[width=7 cm, angle=0]{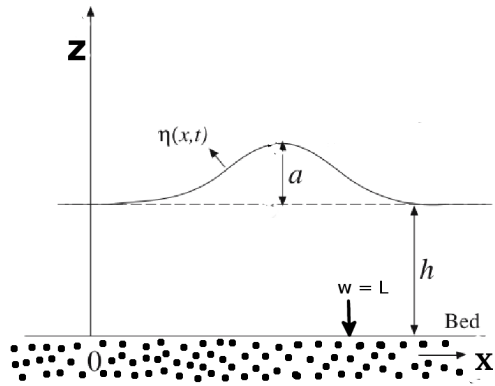}
}

\caption{Solitary wave in shallow water of constant depth with leakage at
the bottom
}
\label{fig:1}
\end{figure}

The purpose of this section is to study the effect of bottom BC with
controlled leakage, designed with a feedback from the surface wave, where the leakage function
  would depend on the wave profile 
and its  spatial derivatives.
As is well known that, the nonlinear free surface gravity waves propagating in a
shallow water in constant depth with the traditional hard bed boundary
condition in the form of solitary waves retain their constant amplitude
profile with a high degree of stability. However, when the boundary
condition is changed to a leakage function dependent on the wave profile
itself, as we find here, the solitary waves propagating on the surface suffer
an amplitude damping along its propagation. Different forms of the leakage
velocity function at the bottom induce different types of damping. Such a
controlled leakage at the bottom may be arranged using a functional feedback
from the profile of the wave appearing on the surface over that location and
at that instant of time. Our motivation for this study is to analyze
different damping effects corresponding to different leakage functions and
identify the case when the damping  would be maximum, which is the
most desirable feature in the present context. 

 In the following
 subsections  we derive  the corresponding free surface wave equation and
investigate  the nature of the    solitary
wave solution with damping caused by different cases of the bottom leakage condition.

\subsection{Surface wave evolution  equation with leakage boundary condition} 

We consider here the
  shallow water nonlinear surface-gravity wave ,
propagating  along
the  positive 
$x$-direction in a  constant  water  depth with
the viscosity and the   surface tension  of the fluid, which is assumed to
be incompressible,  are neglected in
what follows.
We start from the dimensionless  basic   hydrodynamic equations
\cite{Johnson}:
\begin{equation}
 u_t + \epsilon (u u_x + w u_z) = - p_x,  
 \ \ \delta^2[w_t + \epsilon (u w_x + w w_z)] = - p_z,
\label{MomCx1}
\end{equation}
along the x and the z axis, respectively,
which are  reducible from the Euler  equation  in the present  case. 

Here $u, w, p, \eta$
are horizontal and vertical fluid velocity components,  pressure and the
surface wave profile, respectively,
with the  subscripts denoting partial derivatives.
$\epsilon$ is the amplitude parameter defined by $\epsilon
 = \frac{a}{h}$ and $\delta = \frac{h}{l}$ is the shallowness parameter,
expressed through the maximum amplitude $a$, the water depth  $h$  and
 the wavelength $l$ (see FIG. 1).  $\epsilon $ and $\delta $  are 
 natural  parameters supposed to be small, which is 
consistent with the long wave and the shallow water  limit.
 The  continuity equation  of the fluid  
 yields
\begin{equation}
 u_x + w_z = 0.
\label{MassC1}
\end{equation}
 Nonlinear variable boundary conditions, valid at the free
boundary $ z = 1 + \epsilon \eta$, on the other hand,  gives  
\begin{equation}
  p = \eta, \ \ 
 w = \eta_t + \epsilon u \eta_x,
\label{SBC2}
\end{equation}
while we take  the  boundary condition for the vertical component of the
water velocity at the bottom: $z=0$ as
\begin{equation}
 w = - \epsilon \tilde{\alpha} G(\eta, \eta_x,....),
\label{BBC1}
\end{equation}
where $G(\eta, \eta_x,....)$ is assumed, in general, to be an arbitrary function of $\eta$
and its  spatial derivatives and $\alpha'$ is a positive constant with
$\epsilon $ being a small parameter as defined above.
It is important to note here, that usual hard bed scenario with no leakage
one would have $w = 0$ at the bottom whereas in our choice the nontrivial
leakage function $G$ may depend functionally on the surface wave profile
which could be designed through a feedback route, sensing the surface
movement. 
 The leakage is considered
here to be in the $\epsilon $ order.
Note that the negative sign in equation (\ref{BBC1}) appears because the leakage
velocity occurs
along the 
negative  $z$-direction, 
i.e, vertically downward.
 In order to model shallow water solitary waves, there must be an appropriate
balance between nonlinearity and dispersion, i.e, $\delta^2 = O(\epsilon)$
as $\epsilon$ tends to zero.  Thus for any $\delta$, there  exists a region
 in $(x, t)-$ plane with   $\epsilon$ tending to zero, where this balance
remains valid.  This  region of our interest may be  defined by a scaling of
independent variables as $x \rightarrow \frac{\delta}{\sqrt{\epsilon}} x$,
$t \rightarrow \frac{\delta}{\sqrt{\epsilon}} t$ and $w \rightarrow
\frac{\sqrt{\epsilon}}{\delta} w$ for any values of $\epsilon$ and $
\delta$. The set of equations   (\ref{MomCx1}-\ref{BBC1}) thus becomes, \begin{equation}
 u_t + \epsilon (u u_x + w u_z) = - p_x, \ \ 
 \epsilon[w_t + \epsilon (u w_x + w w_z)] = - p_z, \ \ 
 u_x + w_z = 0,
\label{MomCx2}
\end{equation}
together with the boundary conditions
\begin{equation}
  p = \eta,  \ \  
 w = \eta_t + \epsilon u \eta_x, 
\label{SBC22}
\end{equation}
\begin{equation}
 w = -\epsilon \alpha G(\eta, \eta_x,....). \label{BBC2} \end{equation}
 valid at the free surface  and 
 at
the bottom, respectively, 
where $\alpha = \tilde{\alpha} \frac{\delta}{\sqrt{\epsilon}}$, 
with a net outcome of the transformation is to replace $\delta^2$ by
$\epsilon$ in equations (\ref{MomCx1}-\ref{BBC1}).  Introducing a new frame
of reference with stretched time
$ \  \xi = x - t \nonumber,  \ \
\tau = \epsilon t, \ $
we seek an asymptotic solution of the system of equations and boundary conditions in the form
\begin{eqnarray}
 q(\xi,\tau,z;\epsilon) \sim  \sum_{n=0}^{\infty} \epsilon^{n}q_n(\xi, \tau,
z),  \ \
\eta(\xi,\tau;\epsilon) \sim  \sum_{n=0}^{\infty} \epsilon^{n}\eta_n(\xi, \tau) ,
\label{Assexp} 
\end{eqnarray}
where $q$ (and  related  $q_n$) represents each of the functions  $u, w$ and  $p$
for the corresponding expansion.

Now  to deduce the final evolution equation from the set 
of complicated nonlinear equations (\ref{MomCx2}-\ref{BBC2}) involving
several variables, we have to make the asymptotic multi-scale expansions as
explained above. Below, we carry out an explicit  
  order by order calculation to demonstrate the process.
\subsubsection{Result at $\epsilon^0$ order}
At $\epsilon^0$ order the
 above set of equations (\ref{MomCx2})-(\ref{BBC2}) is reduced respectively to the  following set 

\begin{eqnarray}
 u_{0 \xi} = p_{0 \xi}, 
\ \
p_{0z} = 0, \ \
u_{0 \xi} + w_{0 z} = 0 
\end{eqnarray}
\begin{eqnarray}
p_0 = \eta_0, \ \
w_0 = -\eta_{0 \xi},
\label{NVBC0}
\end{eqnarray}
\begin{eqnarray}
w_0 = 0, \label{BBC0}
\end{eqnarray}  
with equation (\ref{NVBC0})  valid at   $z=1$ and (\ref{BBC0})
at  $z=0$.
These equations  lead to the solutions  expressed  through $\eta_0$ as
$ p_0 = \eta_0 , \ \
u_0 = \eta_0, \ \
w_0 = -z \eta_{0 \xi}, \ $
 with the appearance of $\eta$ caused only by the passage of the wave has been
imposed, i.e, $u_0 = 0,$ whenever $\eta_0 = 0$.  \subsubsection{ Result at
$\epsilon$ order } In this order of approximation, two free boundary
conditions at $z = 1 + \epsilon \eta$ are evaluated by performing Taylor
expansions of the functions $u, w, p$ around the point $z=1$.  Consequently
the following set of equations are obtained from
(\ref{MomCx2})-(\ref{BBC2}): \begin{eqnarray}
- u_{1 \xi} + u_{0\tau} + u_{0} u_{0 \xi} + w_0 u_{0z} = -p_{1 \xi},
\ \ p_{1z} = w_{0 \xi}, \ \  
u_{1 \xi} + w_{1 z} = 0 \label{MomCx3}
\end{eqnarray}
with the boundary conditions: 
\begin{eqnarray}
p_1 + \eta_0 p_{0z} = \eta_1, \ \ 
w_1 + \eta_{0} w_{0z} =- \eta_{1 \xi}+ \eta_{0 \tau} + u_0 \eta_{0 \xi},
\end{eqnarray}
  valid at $z=1$.
 We also get
  from the BC at the bottom: $z=0 $, the relation
\begin{eqnarray}
 \ \ w_1 =-\alpha G_0(\eta_0,{\eta_0}_\xi, ...) \label{BBC3}
\end{eqnarray}
where $G_0 $ is the contribution of the  leakage function at $\epsilon^0 $
order.
Using 
the above result,  $w_1$ can be expressed now  as
\begin{equation}
 w_1 = -(\eta_{1\xi} + \eta_{0 \tau} + \eta_0 \eta_{0 \xi} + \frac{1}{2}\eta_{0 \xi \xi \xi}) z
+ \frac{1}{6} z^3 \eta_{0 \xi \xi \xi} - \alpha G_0,
\end{equation}
giving thus all  other functions expressed through 
the fields $\eta_0 $ and $\eta_1  $ only, in this order of approximation.
 Finally 
 eliminating  $ \eta_1$  we  obtain
  the free surface wave equation  as
\begin{equation}
 2 \eta_{0 \tau} + \frac{1}{3} \eta_{0 \xi \xi \xi} + 3 \eta_0 \eta_{0 \xi} + \alpha
G_0 =
0,
\label{FinalE}
\end{equation}
with an additional  term due to the wave profile dependent bottom leakage function 
appearing in   the 
 well known integrable KdV equation\cite{KdV}, which however spoils the integrability
of the system, in general.
 With a scaling of the variables as 
$U = 9\eta_0, T= \tau/6$
 equation (\ref{FinalE})  takes a normalized  form
\begin{equation}
 U_{T}+  U U_{\xi} + U_{\xi \xi \xi} + \beta G_0 = 0,
\label{kdvfor} \end{equation}
where  $\alpha  $ is scaled to $\beta $
and $G_0(U, U_{\xi},..)$ is an  arbitrary smooth function, originating from
the wave profile 
dependent leakage velocity. 
It is fascinating to note, that the condition, we impose for the fluid
velocity at the bottom through a boundary condition with wave profile
dependence makes it way to the nonlinear evolution equation at the surface.

Notice that equation (\ref{kdvfor}) is an extension of the KdV equations with
arbitrary higher nonlinearity, which in general represents a non integrable system.
However  an approximate method due to  
 Bogoliubov and Mitropolsky \cite{Bogoliubov,OttSudan1,OttSudan2} could be applied
here for extracting analytic solutions for the wave equation (\ref{kdvfor})
, in general, in  an implicit form. For explicit analytic solution, one needs to
make suitable choices for function $G_0$. We focus below on some of such
choices with lower order nonlinearities, e.g. $G_0 = U, \ U^2,\  U^3,\ U_{\xi}^2$
though this set, in principle, can be extended further.  
We do not put emphasis on the physical meaning for the individual
forms of the leakage function, since our main motivation is to compare
theoretically the
result of the corresponding wave solutions, to identify the case that would induce
maximum damping of the wave amplitude.
It is intriguing to note, that  similar equations for some of the cases
considered by us  were obtained earlier
 \cite{OttSudan2,Ostrovsky}, though  in completely different physical set-ups.

 In order that this approximation scheme
 to be consistent with the condition for the validity of (\ref{kdvfor}), it is required that
the leakage coefficient $\beta$ should be a small parameter of order
higher than $\epsilon$ as
$1 \gg \beta \gg \epsilon$.

Introducing a  phase coordinate $ \ \ 
 \phi(\xi, T,\beta) =\sqrt{\frac{N(T,\beta)}{12}}(\xi - \frac{1}{3}\int_{0}^{T}N(T,\beta) dT),
\ \ $ through a time-dependent function  $N(T,\beta)$,  assumed to vary slowly with
time, with two different time scales  
$ \ \  t_0 = T, \ \   t_1 = \beta T , $
 we   seek a solution of the wave equation 
following \cite{Bogoliubov}. By expanding  $ U(\phi,\beta,T)$  in small parameter 
$ \beta $ as
\begin{equation}
 U (\phi, \beta, T) = U_0(\phi, t_0, t_1) +  \beta U_1(\phi, t_0) + O(\beta^2),
\label{amplexp2}
\end{equation}
  valid for long times, (as large as $T \sim O(1/\beta)$), we obtain  
using
 (\ref{kdvfor})   an equation
 containing
different powers of $\beta$. Equating coefficients of the same  powers of $\beta$, equations
at different orders are derived, which need to be solved  at each order.

\subsection{Case  $G_0=U $}
We explore 
this case with some details for demonstrating the applicability 
of the Bogoliubov method for solving perturbed KdV equation and 
 for identifying the quantitative trend in the influence of 
 the bottom leakage $G_0 $
on the amplitude of the surface waves.     
Note that the equation obtained in  this case 
 mathematically coencides with the 
dissipation induced evolution considered  in the context of ion-sound waves damped
by ion-neutral collisions   \cite{OttSudan2}.

  Integrating equation (\ref{kdvfor})  for  $G_0=U $,  over the whole range of  $\xi$ we
 can solve  for the total wave amplitude $I(T)=\int_{-\infty}^{\infty}  U d \xi$
and   the total intensity
of the wave  $
P(T)=\int_{-\infty}^{\infty}  U^2 d \xi$ 
to get the explicit expressions as  $I(T) = I(0) \exp(- \beta T)$
and   $P(T) = P(0) \exp(- 2 \beta
T)$,
respectively, where  
 $U(\xi,T)$ and its higher order $\xi$ derivatives are  assumed to vanish
at infinity. 
It is also evident from the exponentially
 decaying nature, that the wave intensity is not conserved in time, confirming that 
the integrability of the perturbed KdV equation (\ref{kdvfor})  in this case 
is lost  
due to the leakage we have considered here.

Since estimating the damping of the  solitary water waves is the main
  concern of our problem, we take the
following relations as  the required initial and boundary conditions:
$ \  U(\phi, 0, \beta) = N_{0} sech^2(\phi), \ \
U (\pm{\infty}, T, \beta) = 0. \ \ $ 
The lowest order equation takes the form
\begin{equation}
 \rho \frac{\partial U_0}{\partial t_0} + \frac{\partial^3 U_0}{\partial \phi^3}
-4 \frac{\partial U_0}{\partial \phi} + \frac{12}{N} U_0 \frac{\partial U_0}{\partial \phi} = 0,
\label{unity}
\end{equation}
where $\rho= \frac{12\sqrt{12}}{N\sqrt{N}}$ with $N(t_1)$ as an  arbitrary function of $t_1$,
except for the initial condition $N(0) = N_0$. 
Solving this equation we obtain
\begin{equation}
 U_{0} (\phi, t_0, t_1) = N(t_1)sech^2(\phi),
\end{equation}
while the $\beta$ order equation takes the form
\begin{equation}
 \frac{\partial U_1}{\partial t_0} + L[U_1] = M[U_0],
\end{equation}
where,
\begin{eqnarray}
 M[U_0] = -\frac{\partial U_0}{\partial t_1} - \frac{\phi}{2N}\frac{\partial U_0}{\partial \phi} \frac{dN}{dt_1}
- U_0, \ \
L[U_1] = \frac{1}{\rho}\frac{\partial^3 U_1}{\partial \phi^3} - \frac{4}{\rho}\frac{\partial U_1}{\partial \phi}
+ \frac{12}{N \rho}\frac{\partial (U_0 U_1)}{\partial \phi}.
\end{eqnarray}
The boundary and initial conditions for $U_1$ are
$ \ \  U_1 (\pm \infty, t_0) = 0,
U_1(\phi, 0) = 0 \ \ $
and it is required that $U_1(\phi, t_0)$ should not
behave secularly with $t_0$.  To eliminate secular behavior of $U_1$ it is
necessary that $M[U_0]$ be orthogonal to all solutions $g(\phi)$ of $L^+[g]
= 0,$ where the function $g(\phi) $ should satisfy $g(\pm \infty) = 0$. 
Here $L^+$ is the operator adjoint to $L$ given by, \begin{equation}
 L^+ = -\frac{1}{\rho}\frac{\partial^3}{\partial \phi^3} +
 \frac{4}{\rho}\frac{\partial}{\partial \phi}
-\frac{12}{\rho} sech^2(\phi)\frac{\partial}{\partial \phi}.
\end{equation}
One  can show,  that the only possible solution of $L^+[g] = 0$,  with $g(\pm \infty) = 0$, 
is in the solitonic form $g(\phi) = sech^2(\phi)$.

Thus from the orthogonality requirement we get 
\begin{equation}
 \int_{-\infty}^{\infty} sech^2(\phi) M[U_0] d\phi = 0,
\label{OrthogC} 
\end{equation}
which yields a simple first order differential equation for $N(t_1)$, the solution of which is
\begin{equation}
 N (t_1) = N(0) exp(-\frac{4 t_1 }{3}), \  \ t_1= \beta T
\label{Decaylaw1}
\end{equation}
for   positive small leakage parameter $\beta$ at large time $T$.
Therefore we obtain the  final result as 
\begin{equation}
 U = N(t_1) sech^2 \phi(\xi,t_1)
 + O(\beta), \ \ \phi(\xi,t_1)
= \sqrt{\frac{N(t_1)}{12}}(\xi + \frac{1}{4\beta }N(t_1)).
\label{finalresult}
\end{equation}
The wave  solution of equation (\ref{kdvfor}) thus obtained for $G_0=U $, shows that the amplitude 
of the solitary wave would  decrease 
with time following (\ref{Decaylaw1}).

Recall that similar dissipative soliton solution was derived  earlier in
many different physical situations \cite{OttSudan2,Ostrovsky}. 
%
%
%

\subsection{ Case $G_0 =  U^2$}

We take up this case for comparison 
 and find that the same Bogoliubov- Mitropolsky method discussed above
is applicable also
in this case with the wave equation taking the form of a
perturbed KdV equation

\begin{equation}
 U_T + U U_{\xi} + U_{\xi \xi \xi} + \beta U^2 = 0.
\label{square}
\end{equation}

Notice, that equation (\ref{kdvfor})
with the  choice for  our leakage velocity function, coincides
formally with the dissipation due to friction at the bottom (Chezy law)
\cite{Ostrovsky}, though for completely different origin. 

Using the same approximation technique, details of which we omit, the decay law
of the solitary wave amplitude for equation (\ref{square}) can  be derived
as
\begin{equation}
 N (T) = \frac{N(0)}{[1 + \frac{16 N(0) \beta}{15} T]}.
\label{decayratesquare}
\end{equation}
Observe, that in comparison with the linear choice of the leakage velocity
the amplitude decay with time becomes weaker in this nonlinear case. To
confirm this trend,which is rather anti-intuitive we take up 
 new cases with enhanced nonlinearity
 and derivatives.

Interestingly, the choice of leakage function as $G_0 = - U_{\xi \xi}$ would
lead to  very similar decay law (\ref{decayratesquare}) and would also coincide
formally with
the effect of magneto-sonic waves damped by electron collisions 
\cite{OttSudan2}.

\subsection{Case $G =  U^3$}
Such a choice of leakage velocity condition with cubic dependence on wave
profile would give rise to the
equation
\begin{equation}
 U_T + U U_{\xi} + U_{\xi \xi \xi} + \beta U^3 = 0,
\label{cube}
\end{equation}
representing  a new perturbed KdV equation, apparently  ignored earlier.
The same approximate treatment leads to the decay law
of the solitary wave amplitude of  (\ref{cube})   as
\begin{equation}
 N = \frac{N(0)}{\sqrt{[1 + \frac{32 N(0)^2 \beta}{35} T]}},
\label{decayratecube}
\end{equation}
decreasing with time as shown in FIG 2.
\begin{figure}[!h]
\centering

{
 \includegraphics[width=7 cm, angle=0]{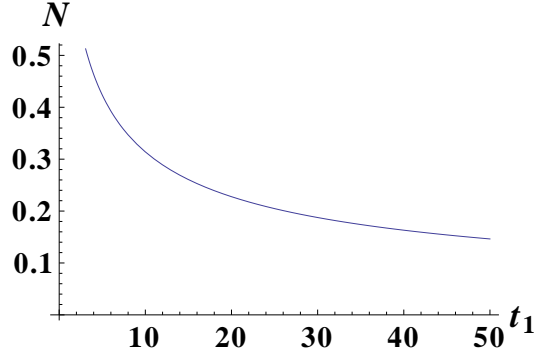}
}

\caption{Plot showing the dependence of the soliton amplitude  $N$ on 
 $t_1$, for the solution (\ref{decayratecube}) with $N(0) = 1.$
 The decaying nature of $N(t_1) $ is explicit.  
}
\label{fig:8}
\end{figure}
Since here for cubic nonlinearity we get the decay rate in
inverse square root power as seen from (\ref{decayratecube}), we notice again that 
 the same trend of the weaker decay of the soliton amplitude
with higher nonlinear dependence of the wave profile on the leakage velocity
function,
continues confirming the anti intuitive trend noticed above.
\subsection{Case $ G = U_{\xi}^2$}
For this choice of the leakage velocity function the perturbed KdV equation reduces to
\begin{equation}
 U_T + U U_{\xi} + U_{\xi \xi \xi} + \beta U_{\xi}^2 = 0,
\label{dersquare}
\end{equation}
apparently not investigated earlier.
Through similar procedure we can derive
the damped solitary wave amplitude of  (\ref{dersquare}) as
\begin{equation}
 N = \frac{N(0)}{\sqrt{[1 + \frac{8 N(0)^2 \beta}{45} T]}},
\label{decayratedersquare}
\end{equation}
which is graphically represented in FIG 3.
\begin{figure}[!h]
\centering

{
 \includegraphics[width=7 cm, angle=0]{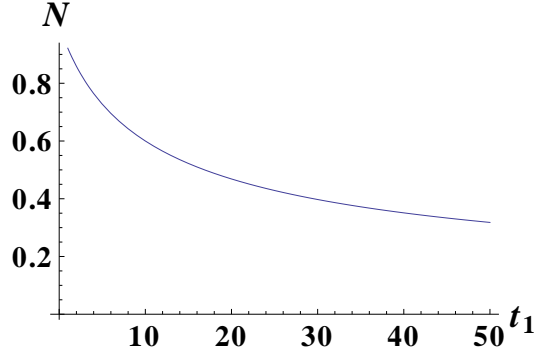}
}

\caption{The solitonic wave amplitude  $N(t_1)$ (\ref{decayratedersquare})
 as decays with time $t_1$ for 
   $N(0) = 1$.
}
\label{fig:9}
\end{figure}

Comparing (\ref{decayratesquare}) with (\ref{decayratedersquare}) we may
conclude, that the increase of nonlinearity as well as derivatives, 
of the wave profile in the leakage velocity function weakens the decay rate of
the solitonic amplitude.
Analyzing the above results for linear and nonlinear choices of $G_0 $,
  we may conclude that the leakage with the linear
dependence on the profile $G_0 = U$ is the optimal one capable of inducing maximum decay
rate on the soliton amplitude as exponential functions, compared to all other cases
considered here. Therefore in the
next sections we take up this particular case, being the most desirable one, for controlling the surging waves in a
decreasing depth scenario.


   
\section {Effect of leakage BC on  nonlinear shallow water surface wave in  variable depth
bathymetry}

\begin{figure}[!h]
\centering

{
 \includegraphics[width=7 cm, angle=0]{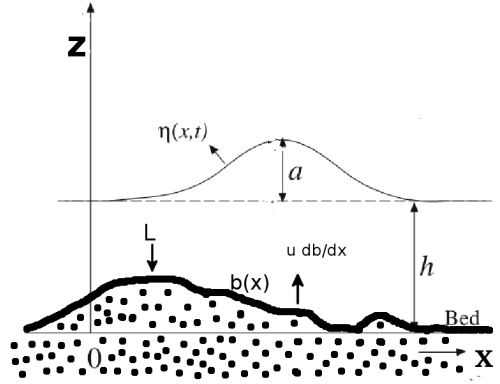}
}

\caption{Solitary wave in a shallow water of slowly  varying depth with
leakage at the bottom  
}
\label{fig:4}
\end{figure}

Propagation of nonlinear shallow water unidirectional  waves
over variable depth topography has been studied intensively with 
rich results
\cite{Johnson,Ostrovsky,Kundu,Pelinovsky}.

 It is known that
the slowly variable depth in comparison to the evolution scale of the wave,
can lead to the upsurging wave amplitude, for decreasing depth, which occurs
when the wave approaches to the shore.
In this section we intend to focus on
such a situation due to its potentially hazardous consequences and
look for its possible regulation through bottom leakage.
Since in the previous section we have identified the maximum damping effect
of surface waves for leakage velocity function depending linearly on the wave
profile,
we will apply this particular leakage condition to achieve maximal damping effect.
Therefore we take up the problem of
 nonlinear wave propagation  
over shallow water of slowly varying depth in the framework of KdV
equation, together with a nontrivial leakage condition
at the bottom with a leakage function proportional to the surface wave
profile, sensed through a feedback mechanism.

This problem targeted towards controlling the surging waves due to
decreasing depth bathymetry, has not received the needed
attention.


\subsection{Derivation of nonlinear surface wave evolution equation with slowly variable
depth under bottom boundary leakage condition} 

Under this physical situation one has to start with the same
 basic dimensionless hydrodynamic equations considered in the previous
section as:
\begin{eqnarray}
 u_t + \epsilon (u u_x + w u_z) = - p_x, \ \
 \epsilon[w_t + \epsilon (u w_x + w w_z)] = - p_z, \ \
 u_x + w_z = 0,
\label{MomCxv}
\end{eqnarray}  
together with the surface boundary conditions 
 $ \ \  p = \eta, \ \ 
 w = \eta_t + \epsilon u \eta_x$
valid at $ z = 1 + \epsilon \eta$.  
However, the effect of variable depth
and the leakage condition enter through a more general boundary condition
at the bottom, varying as $z = b(x)$:
\begin{equation}
 w = u \frac{d b}{ d x} - \epsilon  g(\epsilon x) G(\eta, \eta_x,....).
\label{BBCv}
\end{equation}
 Note that in comparison with the previous case (\ref{BBC1}) together with the
variable depth function an additional leakage function $g(\epsilon x)$
 independent of the wave profile  $\eta$  appears with $G$ similar to the
feedback leakage function as considered in the previous section.
The bathymetry function
$b$  is assumed to depend on the small parameter $\epsilon$, such that $b (x) = B(\epsilon x)$.
As we have identified in previous section, we assume $G = \eta$ to get the maximum benefit
of damping due to leakage.
 For detailed investigation we introduce a new set of variables
\begin{eqnarray}
 \xi = \frac{1}{\epsilon} \chi(X)  - t,   \ \
X = \epsilon x,
\label{chiF}\end{eqnarray}
where $\chi(X)$ will be determined later in equation (\ref{chiFx}).
For solving the above set of equations we would represent 
the asymptotic solutions as we have used earlier.

We stress again that the hydrodynamic equations involved here are the
 same as those used in the previous
section in dealing with the constant depth problem, except the crucial BC at
the bottom. 
\subsubsection{Result at $\epsilon^0$ order}
 
At $\epsilon^0$ order, the above equations are reduced to
\begin{eqnarray}
 u_{0 \xi} = \chi' p_{0 \xi}, \ \ 
p_{0z} = 0, \ \
\chi' u_{0 \xi} + w_{0 z} = 0 
, \label{bulk2}\end{eqnarray}
together with the boundary conditions \ \
$p_0 = \eta_0, \ \
w_0 = -\eta_{0 \xi},$\ \
valid at the surface and
$w_0 = 0,$\ \
 at the variable bottom  
 $z= B(X)$.

Using the above bulk equations and the boundary conditions  
we obtain
\begin{eqnarray}
 p_0 = \eta_0, \ \ 
u_0 = \chi' \eta_0,  \ \
w_0 = \chi'^2 \eta_{0 \xi} (B -z), \ \
\chi'^2 = \frac{1}{D(X)},
\end{eqnarray}
where $D(X) = 1 - B(X)$ and $\chi'$ is the derivative of $\chi$ with respect to
$X$.
 $\chi $ can be solved explicitly through the bathymetry function for the
right moving wave as 
\begin{equation}
 \chi (X) = \int_0^{X} \frac{d X_1}{\sqrt{D(X_1)}}.
\label{chiFx}\end{equation}

\subsubsection{$\epsilon$ order approximation}
In next order approximation we obtain the set of equations
\begin{eqnarray}
- u_{1 \xi} + \chi' u_{0} u_{0 \xi} + w_0 u_{0z} = -\chi'p_{1 \xi}- p_{0 X},\ \
p_{1z} = w_{0 \xi}, \ \
\chi'u_{1 \xi} + u_{0 X} +w_{1 z} = 0 \ \
\label{MomCx4}
\end{eqnarray}
together with the surface boundary conditions 
\begin{eqnarray}
p_1  = \eta_1, \ \
w_1 + \eta_{0} w_{0z} =- \eta_{1 \xi} + u_0 \chi' \eta_{0 \xi},\ \
\label{SBC24}
\end{eqnarray}
and the condition 
\begin{eqnarray}
w_1 = u_0 B'(X) - g(X) \eta_0, \label{BBC4}
\end{eqnarray}   valid at the variable bottom with  $B'(X)$ denoting  derivative 
in  $X$. Our aim is to express other field variables only through the wave
functions $\eta_0$ and $\eta_1 $  as
\begin{equation}
 p_1 =  \eta_1 + \frac{1}{D} \eta_{0 \xi \xi} [\frac{1}{2}(1 - z^2) + B(z -
1)]
\end{equation}
and
\begin{eqnarray}
 w_1 = (\frac{B'}{\sqrt{D}} - g)\eta_0 + \frac{(B-z)}{\sqrt{D}}\eta_{0X}
+ (B-z)(\frac{\eta_0}{\sqrt{D}})_X + \frac{(B-z)}{D}\eta_{1 \xi} + \frac{(B - z)}{D^2} \eta_0 \eta_{0 \xi}\nonumber\\
- \frac{\eta_{0 \xi \xi \xi}}{D^2}[B(\frac{z^2}{2}- z)+ \frac{(z - \frac{z^3}{3})}{2}
- \frac{B^3}{3} + B^2 - \frac{B}{2}]. \label{w1}
\end{eqnarray}

Using the above expressions 
  we can finally derive the 
 surface wave evolution equation  
\begin{equation}
 2 \sqrt{D} \eta_{0X} + \frac{3}{D}\eta_0 \eta_{0 \xi} + (\frac{D'}{2
\sqrt{D}}+ g)\eta_0
+ \frac{D}{3} \eta_{0 \xi \xi \xi} = 0.
\label{vckdv}
\end{equation}
Note that this variable coefficient KdV
 equation contains explicitly the bathymetry function $D(X)$ linked to the
variable depth
as well as the function $g(X)$ related to the leakage at the bottom.
This variable coefficient KdV equation containing the combined effect of
variable depth and the leakage is an important result we have derived here.
Different types of variable coefficient KdV like equations were studied earlier
for analyzing the possible solutions both in one \cite{PRE84,PRE83,PLAsol1,PLAsol2}
and two dimensions \cite{PLA2dkdv,PREablowitz}. 
  
\subsection{Nature of the solitary wave solution}
It is evident that in the absence of the leakage $(g = 0)$, our equation (\ref{vckdv})
would reduce to the KdV equation
 with variable depth \cite{Johnson,Ostrovsky,Kundu}:
\begin{equation}
 2 \sqrt{D} \eta_{0X} + \frac{3}{D}\eta_0 \eta_{0 \xi} + (\frac{D'}{2 \sqrt{D}})\eta_0
+ \frac{D}{3} \eta_{0 \xi \xi \xi} = 0.
\label{kdvJohnson}
\end{equation}

When the depth variation occurs in a scale slower than the evolution
scale  of the wave, the solitary wave solution of equation
(\ref{kdvJohnson}),as is wellknown,  can be expressed as
 an approximate solution 
\begin{eqnarray}
 \eta_0 = \frac{A_0}{D}sech^2 {[\sqrt{\frac{3 A_0}{4 D^3}}(\xi -
\frac{D^{-(\frac{5}{2})} A_0 X}{2})]},
\label{surging}
\end{eqnarray}
as given in \cite{Johnson}.
Here $A_0$ is the amplitude of the wave for constant depth ($D = 1$). 
It is clearly seen that the amplitude of the solitary wave increases as $D$ decreases
 i.e. the water becomes
shallower,  showing that 
such waves would approach the shore with surging amplitude. Note that for
exponentially decreasing depth $D$ the growing of wave amplitude will also
be exponential. This particular case will be considered in more details in the next
section.

It is intriguing to note that for variable bathymetry with uneven depth,
irregular depth or periodic topography in place of growing amplitude one gets a
damping wave amplitude as explained in \cite{Pelinovsky}. We will be
concerned however with the surging waves caused by a smoothly decreasing depth
due to their hazardous effects.

Now we will analyze the solution of
equation (\ref{vckdv}) with nontrivial boundary leakage, rewriting it
 in a more general form 
\begin{equation}
 a(X) \eta_{0X} + b(X) \eta_0 \eta_{0 \xi} + c(X) \eta_0 + d(X) \eta_{0 \xi \xi
\xi}=0,
\label{vkdvabcd}
\end{equation}
where we have denoted $a(X)= 2 \sqrt{D}, b(X) = \frac{3}{D}, c(X) = (\frac{D'}{2 \sqrt{D}}+ g)$
and $d(X) = \frac{D}{3} $.
Dividing (\ref{vkdvabcd}) by $d(X)$ and defining $\eta_0 = \frac{U}{b_1}$
where $a_1 = \frac{a}{d}, b_1 = \frac{b}{d}$ and $c_1 = \frac{c}{d}$,
respectively, the equation
(\ref{vkdvabcd}) can be transformed to
\begin{equation}
 a_1 U_X + U U_{\xi} + U_{\xi \xi \xi} + (c_1 - a_1 \frac{b_{1X}}{b_1})U =
0,
\label{MainE} 
\end{equation}
which in general cannot be solved exactly.
However, we may notice, that for a finer balance tuned between the variable depth
bathymetry and the controlled leakage velocity function giving the condition
\begin{equation}
  g = - \frac{9 D'}{2 \sqrt{D}},
\label{gD}\end{equation}
the last term of (\ref{MainE}) vanishes reducing the equation to a more
 simple form of 
 variable coefficient  KdV
equation 
\begin{equation}
 a_1 U_X + U U_{\xi} + U_{\xi \xi \xi} = 0, \label{xKDV}
\end{equation}
where $a_1 = \frac{a(X)}{d(X)}$.
It is interesting to note, that
the tuning condition (\ref{gD}) relating the leakage function with the 
bathymetry function is exactly same as the  solvability condition used in 
\cite {PRE83} for obtaining analytic solutions of a general variable
coefficient KdV equation, considered in a formal mathematical setting.

 Defining  a new coordinate  $T = \int \frac{\sqrt{D(X)}}{6} dX $ 
equation (\ref{xKDV}) can be transformed  into the standard constant coefficient
KdV equation
\begin{equation}
  U_T + U U_{\xi} + U_{\xi \xi \xi} = 0,
\end{equation}
admitting the well known   solitary wave solution
 $U = N_0 sech^2[\sqrt{\frac{N_0}{12}}(\xi - \frac{N_0}{3} \int
\frac{\sqrt{D}}{6}dX)].$
Expressing in terms of the original field variable we get finally  the wave
solution  
\begin{equation}
 \eta_0 = \frac{D^2}{9}N_0 sech^2[\sqrt{\frac{N_0}{12}}(\xi - V(X)],
 \ \ V(X)= \frac{N_0}{3} \int \frac{\sqrt{D}}
{6} dX
\label{exactsolution}
\end{equation}
with the depth function $D(X)$ and leakage velocity function $g(X)$
are tuned as (\ref{gD}).
Note that for decreasing depth $D$, which without leakage would make the wave
amplitude to surge as in (\ref{surging}), due to the controlled tuning of the leakage
the resultant solitonic wave function  would suffer a damping of its amplitude as evident
from (\ref{exactsolution}). Moreover the solitonic wave flattens down with a change in its
velocity along its propagation (see FIG 5).
Thus we have achieved 
control over a surging wave approaching to the shore by inducing
combination of feedback and a controlled tuning of the 
the leakage at the bottom. 
   
\begin{figure}[!h]
\centering

{
 \includegraphics[width=7 cm, angle=0]{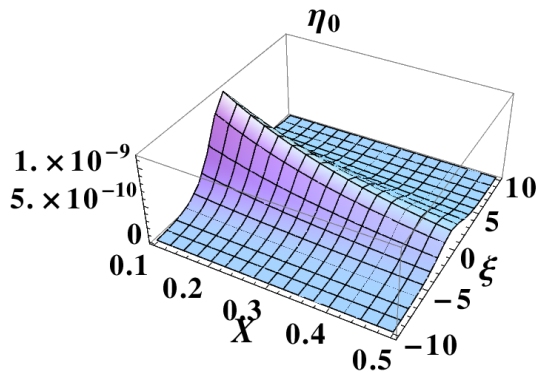}
}

\caption{3D  plot of the solitary wave solution
 (\ref{exactsolution}) in the $(\xi, X) $ plane. For demonstrating the
nature of the solution, we have assumed $N(0) = 1$
 $\alpha=0.1 $, $g = \exp{[-X]}$, showing  exponential damping of the wave amplitude
 with a change in its width and velocity along its
 propagation.  }
\label{fig:5}
\end{figure}
\section{ application of the exact result to  real near shore bathymetry}

In the previous sections, we have first discussed the effect of wave profile dependent leakage
to the solitary wave amplitude at constant water depth. Applying similar mathematical procedure to a
slowly changing bathymetry, we have derived next a variable coefficient KdV equation containing terms due 
to both leakage and variable depth. Though in general such equations are non integrable,
a finer balance between the leakage  and variable depth function
miraculously solves the equation exactly, giving a solitary wave like solution. Its amplitude, which
without leakage would increase giving surging effects, decreases
as the wave moves towards the shallower region. These theoretical findings of exact nature 
with an intension to control near shore surging waves, by creating
 artificial leakage, would gain ground
when it is  implemented to a real sea shore bathymetry. Therefore, in this section we 
apply  previously obtained exact results to a near shore bathymetry in order to see the effectiveness 
of our findings.

One should remember the fact that according to the estimates of the United Nations in 1992,
 more than half of the population
lives within 60 km of the shoreline. Urbanization and rapid growth of coastal cities have
also been dominant
population trends over the last few decades, leading to the development of numerous mega cities in all
coastal regions around the world.

 Our study region is the coastal zone of Chennai district
of the Tamil Nadu state, in southeast coast of India which was one of the worst affected areas during
2004 Indian Ocean tsunami. A Coastal Vulnerability Index was developed for this region 
\cite{ArunKumar} using eight relative risk variables including near shore bathymetry to know the high
and low vulnerable areas. According to one of those risk variables, bathymetry at about 29.11
km of coastline in that area has a high risk rating having high vulnerability
,while about 18.55 km of coastline has medium risk rating
and about 10.54 km shows low risk rating, which are displayed in FIG 6.
\begin{figure}[!h]
\centering

{
 \includegraphics[width=10 cm, angle=0]{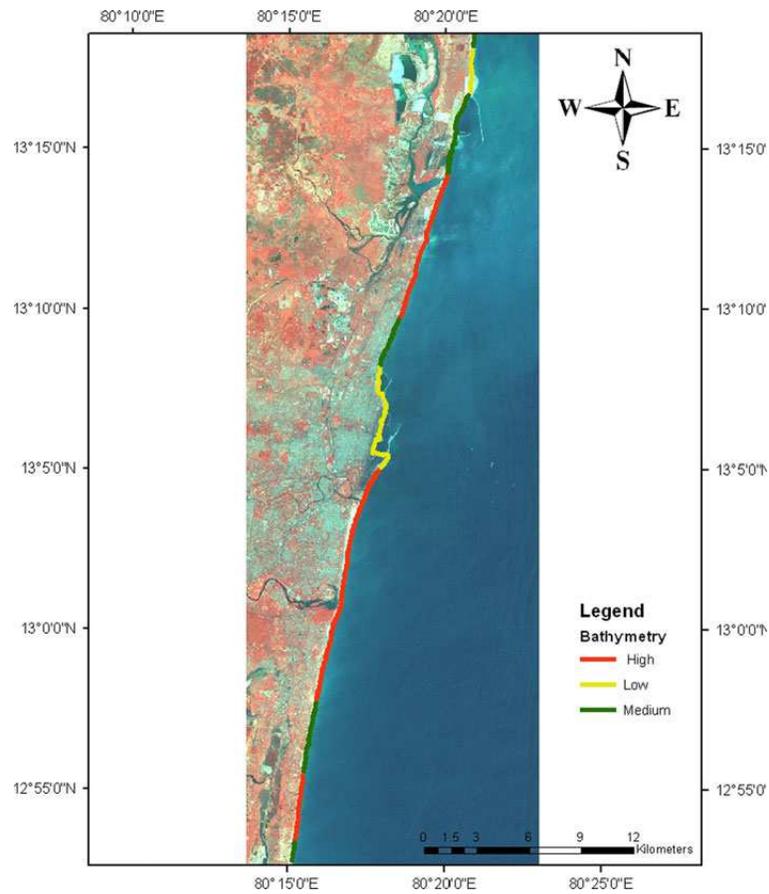}
}

\caption{ Risk zones of Chennai coastline bathymetry ,taken from \cite{ArunKumar} with permission}
\label{fig:5}
\end{figure}
 
The depth contour of Chennai coastline, which is constructed from the 
Naval Hydrographic Charts for 2002, is also given in \cite{ArunKumar} and is displayed in FIG 7.
Now to implement our exact results on this coastline, we chose one of the high risk points
(N $13^\circ$ $10.5'$ - E $80^\circ$ $18.75'$)  
, which is denoted by the red line in FIG 6.
\begin{figure}[!h]
\centering

{
 \includegraphics[width=8 cm, angle=0]{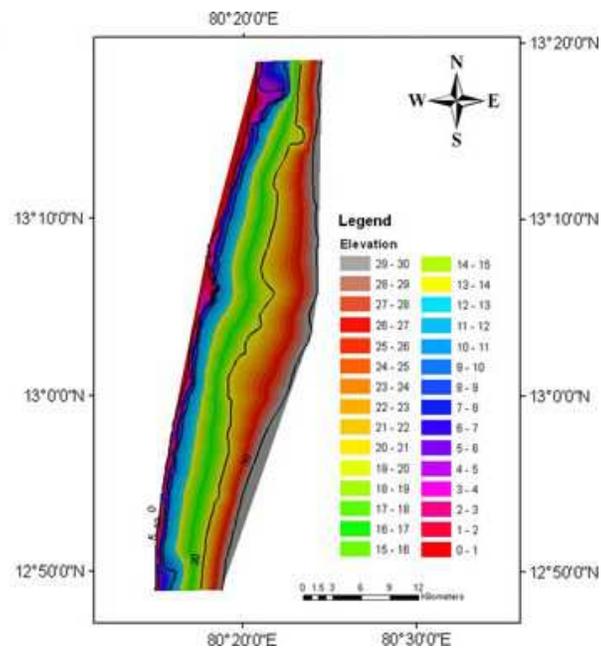}
}

\caption{ Depth contour of the Chennai coastline,taken from \cite{ArunKumar} with permission}
\label{fig:5}
\end{figure}

 We have drawn the near shore bathymetry following the depth contour 
(FIG.7) of this shoreline point along the latitude which is given as FIG.8.This
 diagram shows that at the near shore region, the depth function flattens down
denoting a slow variation along $X$. Hence the soliton gets enough time to evolve to give the surging
effects. Note also that, the variation along  $X$ is in km whereas variation along $D(X)$ is
in meter. Hence the depth function is very slowly varying which is consistent
with our theoretical assumptions.

Note that in the absence of the leakage at the bottom,  the solitary wave amplitude
would increase following (\ref{surging}) with the amplitude as
\begin{equation}
A_1 = \frac{A_0}{D}
\label{A1}.
\end{equation}
We see from FIG.8 that as the wave approaches the near shore region, the depth function
flattens out and therefore the soliton amplitude $A_1$  develops rapidly
to give surging effects.
  Now if at a certain position in the near
shore bathymetry, an artificial leakage following our theoretical findings (\ref{exactsolution}), is turned on then
the amplitude  would decrease as $A_2 = \frac{N_0 D^2}{9}$, where $N_0$ is a free constant
 The effectiveness  of the amplitude decay of the solitary waves by the leakage
would be stronger, if the leakage starts at a longer distance away from the shore.
Note that the amplitude starts growing rapidly at 1.2 km away from the shore, from where the depth function
starts flattening. 

Therefore, if a solitary wave of amplitude of nearly 1 meter starts approaching towards the shore from
around 10.5 km, then it would ultimately  grow to a surging wave of amplitude $\sim$ 30 meter at the coast.
It is obvious that such a huge wave will produce devastating effects on coastal habitation
and costly installations. 

However if we implement now an artificial leakage based feedback method 
linked to the surface wave profile
as discussed in the previous section with exact result
(\ref{exactsolution})  the surging amplitude would decrease when propagating towards the shore.
with damping amplitude given as (\ref{exactsolution})

\begin{equation}
   A_2 = \frac{N_0 D^2}{9} 
\label{A2},
\end{equation}
where $N_0$ is a free constant which is chosen following the actual physical condition as described earlier. One checks that
if the leakage is implemented in a region of 0.9 km from the shore (at the point $Q_1$ in FIG.9),  the wave amplitude 
of 1 meter which would otherwise increase to 30 meter without any leakage (denoted by point $D$ in FIG.9),
would decrease to  an amplitude of $\sim$ 1.23 meter ( denoted by point $A$ in FIG.9),
where we have chosen $N_0 = 11.60$. 

If the leakage installation is implemented from a nearer point from the shore, the wave amplitude decrease
would also be less which is also displayed in FIG.9. For optimal estimation however,
 the cost effectiveness and the concrete
requirements should be taken into account in deciding the range of such proposed installations. The main
emphasis should possibly be on the protection of sensitive installations
like nuclear reactors at the sea coast against the danger of tsunami like waves. 
 The 
options known for the protection of the Chennai coast  area 
are dune afforestation, mangrove restoration and management, periodic beach nourishment
and building seawalls and groins etc. Our control mechanism for the possible management of the potentially
hazardous near shore waves, proposed here, could be a new option, which may be implemented
 only in  limited strategic areas surrounding costly installations, for reducing the intensity of the 
approaching wave to a safer limit.

\begin{figure}[!h]
\centering

{
 \includegraphics[width=7 cm, angle=0]{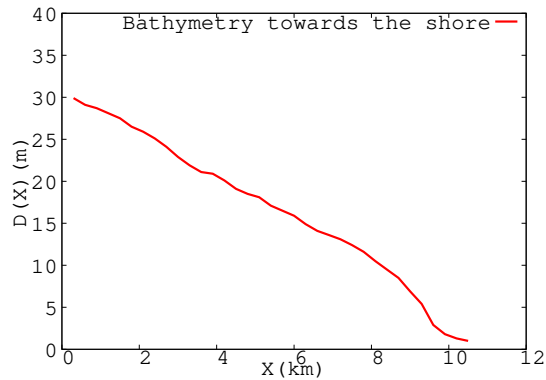}
}

\caption{Bathymetry towards the shore of the shoreline point ( N $13^\circ$ $10.5'$ - E $80^\circ$ $18.75'$)  }
\label{fig:5}
\end{figure}

\begin{figure}[!h]
\centering

{
 \includegraphics[width=10 cm, angle=0]{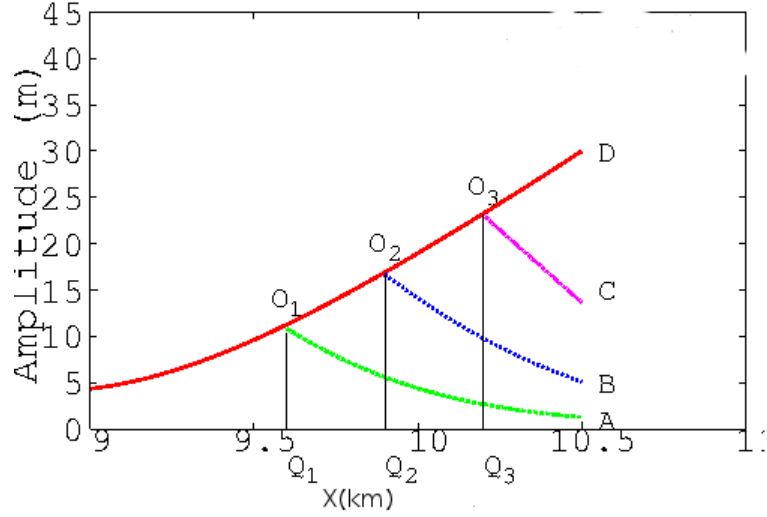}
}

\caption{Surging amplitude $A_1$ without leakage moving towards the shoreline point 
 ( N $13^\circ$ $10.5'$ - E $80^\circ$ $18.75'$) and growing upto the point D (30 m) following eq.(\ref{A1}). 
 Figure also demonstrates  the damping of the amplitude $A_2$ due to leakage following eq. (\ref{A2}) .
 Installations of the leakage starting from different points to the shore
$Q_1$ (9.6 km), $Q_2$ (9.9 km), and $Q_3 $ (10.2 km), would damp the amplitude
$A_2$  to different values ( $A$ (1.23 m), $B$ (5.15m) and $C$ (13.5m) respectively).     $N_0$, a free constant appearing in eq. (\ref{A2})
is chosen as 11.60, 46.29 and 122.89 respectively at these points.  It is evident that,  the further the leakage is   from the shore,
the more the decay of the amplitude. }
\label{fig:5}
\end{figure}

The same methodology can be applied to another high risk point ( N $13^\circ$ $0'$ - E $80^\circ$ $16.2'$)
 at the shoreline, the near shore bathymetry of which is shown in FIG 10. The increase of amplitude 
without leakage, and its damping due to leakage is explicit in FIG.11.

\begin{figure}[!h]
\centering

{
 \includegraphics[width=7 cm, angle=0]{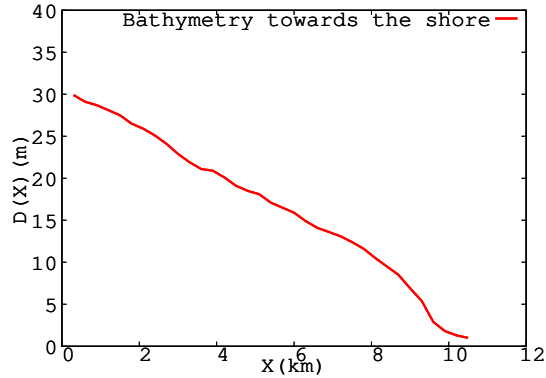}
}

\caption{Bathymetry towards the shore of the shoreline point ( N $13^\circ$ $0'$ - E $80^\circ$ $16.2'$) }
\label{fig:5}
\end{figure}

\begin{figure}[!h]
\centering

{
 \includegraphics[width=10 cm, angle=0]{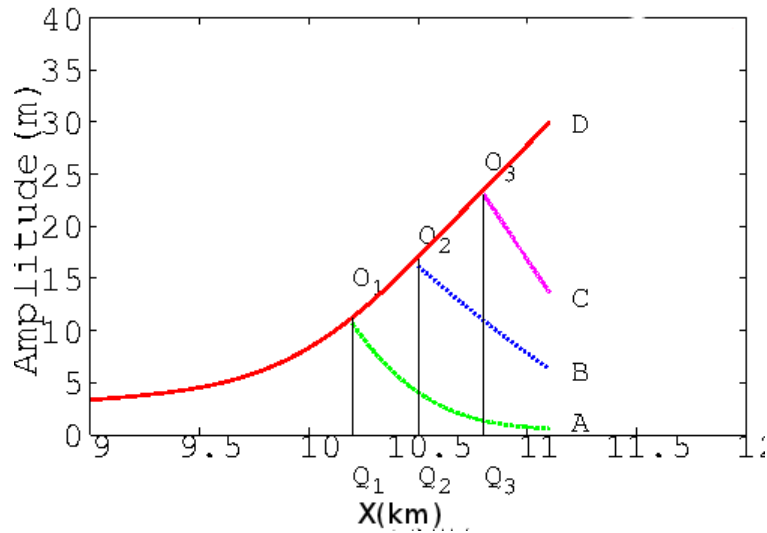}
}

\caption{ Surging amplitude $A_1$ without leakage moving towards the shoreline point 
 ( N $13^\circ$ $10.5'$ - E $80^\circ$ $18.75'$) and growing upto the point D (30 m) following eq.(\ref{A1}). 
 Figure also demonstrates  the damping of the amplitude $A_2$ due to leakage following eq. (\ref{A2}) .
 Installations of the leakage starting from different points
$Q_1$(10.2 km), $Q_2$ (10.5 km) and $Q_3$ (10.8 km), would damp the amplitude
$A_2$  to different values ( $A$ (0.53m), $B$(6.24m) and $C$(13.57m) respectively).     $N_0$, a free constant appearing in eq. (\ref{A2})
is chosen as 5.44, 56.91 and 122.89 respectively at these points.  It is evident that,  the further the leakage is   from the shore,
the more the decay of the amplitude.
}
\label{fig:5}
\end{figure}

\section{Concluding remarks}
The focus of our investigation is  an innovative possibility of
 controlling the intensity of near shore surging waves including tsunamis
and bore waves, by inducing damping effect through a specially designed
leakage mechanism at the water bed.

The majority of the earlier studies, concentrated on the damping  of
 the waves occurring due to natural effects like viscosity, bottom roughness,
sand porosity etc. In contrast, our main 
motivation here is to analyze the
 impact of artificially created bottom boundary condition on the swelling wave
approaching  the shore,
with an aim to reduce the hazardous effect
of such near shore wave phenomenon.

Our crucial observation is, that the surging of approaching waves caused by
decreasing water depth bathymetry may be thought of to be triggered by
effective vertical fluid flow proportional to the gradient of the depth
profile, acting as a virtual {\it source} emerging from the bottom. Our key
idea for controlling the growing amplitude of the surface wave is to counter
this source by an effective {\it sink} through such leakage mechanism
creating a downward fluid velocity.

We have considered the propagation of
an unidirectional, shallow water, nonlinear free surface gravity wave
based on the basic hydrodynamic equations at the shallow water regime and
identified first, that a feedback leakage function at the bottom, dependent
linearly on the surface wave profile, could induce maximum desirable damping
effect on the amplitude of the surface wave. This knowledge is then applied
to the problem of regulating the surging 
solitary waves propagating towards the shore, due to the slowly decreasing
depth. The corresponding evolution equation for the combined effect of
leakage and the variable bathymetry turns out to be in the form of a
variable depth KdV equation, different from the variable coefficient KdV
equation obtained earlier. Though in general this is a non-integrable system,
we have found, that for a controlled tuning between the topography and leakage
velocity function, the equation becomes exactly solvable, allowing solitary
wave solutions with damping amplitude. 

A strong point of our result is its exact nature, which allows one to access
precise and finer effects and make more accurate predictions. We have
applied the result obtained to real data from the bathymetry map of the
tsunami prone near shore regions on the Bay of Bengal in India and tested
the implications, range and predictions of our theoretical result.
As shown by the real bathymetry , the more extensive installations
starting from a further distance into the sea would result to a more
effective control of the incoming surging waves. However, the cost
effectiveness and the concrete requirements should be taken into account in
deciding the range of such proposed installations. The main emphasis should
possibly be on the protection of sensitive installations like nuclear
reactors at the sea cost against the danger of tsunami like waves.
Therefore the control mechanism for the possible management of the
potentially hazardous near shore waves, proposed here, may be implemented
only in limited strategic areas surrounding costly installations, for
reducing the intensity of the approaching wave to a safer limit.

We have studied also various possible extensions of the leakage boundary
conditions and their corresponding effects in modifying the nature of the
surging solitary waves which might be of practical importance in different
other situations.(This material is included as the appendix.)   


\section{Appendix: Extension of boundary leakage condition with variable bathymetry }
Though we have achieved our major goals in taming the surging waves
as reported in the main text, we
consider below few extensions of this result for understanding the effect of
bottom boundary leakage condition on the surface wave solution, which might
be of applicable interest in other physical situation. In particular ,we
have investigated

{\bf A}) Leakage function at the bottom with a combination of both wave profile
dependent and independent functions,

{\bf B}) Leakage condition linked to effective zero fluid velocity at the
bottom with the specific bathymetry profile.

{\bf C})Leakage function related to time.
 
All the studies yielding analytic result of different nature though all of them
having the effect of amplitude damping of the waves, surging otherwise due to
decreasing depth bathymetry.  

\subsection{Leakage function at the bottom with a combination of both wave profile
dependent and independent functions}
In our previous paper \cite{mypaper}, we considered the leakage function to be 
independent of the wave profile that yielded  a forced KdV like equation 
as the surface wave equation. Its solitary wave solution exhibits phase modification
leading its velocity to change whereas the amplitude remaining constant.
In order to explore the effect of the bottom leakage  on the solitary wave amplitude
we have considered in the main text, the leakage function to be dependent on the free surface wave profile
which exhibited damping of amplitude.
 
Now in this section, we have extended the problem such that the leakage velocity at the bottom depends  both on  
 the  wave profile dependent and  independent functions as 
\begin{equation}
 w = u \frac{d b}{ d x} - \epsilon  g(\epsilon x) G(\eta, \eta_x,....) + \epsilon C(X).
\label{BBCg}
\end{equation}
on $z = B$. Here the second term in (\ref{BBCg}) is the wave profile dependent term whereas the third one
is the wave profile independent term.
 
As we have mentioned we assume $G = \eta$ to get the maximum benefit
of damping due to leakage.
 After a bit of mathematical calculations 
  we can finally derive the 
 surface wave evolution equation  
\begin{equation}
 2 \sqrt{D} \eta_{0X} + \frac{3}{D}\eta_0 \eta_{0 \xi} + (\frac{D'}{2
\sqrt{D}}+ g)\eta_0
+ \frac{D}{3} \eta_{0 \xi \xi \xi} = -C(X).
\label{vckdvg}
\end{equation}
Note that this variable coefficient KdV
 equation contains explicitly the bathymetry function $D(X)$ linked to the
variable depth
as well as the function $g(X)$ and $C(X)$ related to the leakage at the bottom.

Now after applying  the same balancing condition (\ref{gD}) the equation can be transformed into 
\begin{equation}
 a_1 U_X + U U_{\xi} + U_{\xi \xi \xi}  = -E_1
\label{MainE2} 
\end{equation}
where we have denoted $a_1(X) = \frac{6}{\sqrt{D}} $, $E_1= \frac{27 C(X)}{D^3}$ and $\eta_0 = \frac{U}{b_1}$.

Defining  a new coordinate  $T = \int \frac{\sqrt{D(X)}}{6} dX $ 
equation (\ref{MainE2}) can be transformed  into the standard constant coefficient
KdV equation with a forcing term
\begin{equation}
  U_T + U U_{\xi} + U_{\xi \xi \xi} = E_1(T),
\end{equation}
admitting the well known   solitary wave solution
 $U = N_0 sech^2[\sqrt{\frac{N_0}{12}}(\xi - \frac{N_0}{3} \int
\frac{\sqrt{D}}{6}dX) - f(T)] - \int E_1 dT  $.

Expressing in terms of the original field variable we get finally  the wave
solution  
\begin{equation}
 \eta_0 = (D^2/9) [N_0 sech^2\{\sqrt{\frac{N_0}{12}} (\xi - \frac{N_0}{3} \int
\frac{\sqrt{D}}{6}dX - f(T))\} - \int E_1 dT]
\label{gen}
\end{equation}
where $\frac{\partial^2f(T)}{\partial T^2} = -E_1$.

Note that if we neglect the wave profile independent part $C(X)$, then automatically we get
$C_1 = E_1 = F  = 0$ and $f = $ constant. Thus the solution (\ref{gen}) converges to the solution of earlier case
 (\ref{exactsolution}).

 \subsection{Balancing through effective hard bottom condition with leakage
giving exact
result}
Here we stick to a particular choice of decreasing bathymetry $D = \exp{(-\sigma X)}$, 
for the wave
approaching to the shore. Such solitary waves
without any leakage condition would result to an exponentially surging waves
carrying potential hazards. Our aim here would be to control such wave
 through bottom leakage condition inducing necessary damping. For this
purpose we
 consider a different balancing effect of the
leakage condition,
 obtained from an effective hard bottom condition
amounting to the vertical fluid velocity at the water bed $w$ to be zero.
This leads at the leading order to
 $w_0 = 0$,  $ w_1 = u_0 B'(X) - g(X) \eta_0 = 0$, at $z = B$, which gives
a new balance between the leakage and the variable depth function as
$ g = -\frac{D'}{\sqrt{D}}$ at $z = B$.
For this effective hard bottom condition, we 
follow again similar mathematical procedure as presented in the previous
section, which leads to
the surface wave evolution equation   
\begin{equation}
  2 \sqrt{D} \eta_{0X} + \frac{3}{D}\eta_0 \eta_{0 \xi} - \frac{D'}{2 \sqrt{D}} \eta_0
+ \frac{D}{3} \eta_{0 \xi \xi \xi} = 0,\ \ D = D(X).
\label{special}
\end{equation}

Note that this variable coefficient KdV equation is different from the variable
bathymetry  equation (\ref{kdvJohnson}) obtained earlier \cite{Johnson}. 
As such this equation is also difficult to solve analytically. However
interestingly for a special choice of bathymetry function 
 $D = \exp{(-\sigma X)}$,  with
 $D$ decreasing with the increase of $X$, which is consistent with
the wave propagating towards shallower region, we can find an exact wave solution
for equation (\ref{special}).

Dividing equation (\ref{special}) by $\frac {D} {3}$ and
 redefining the field as $\eta_0 = \frac{D^2}{9} H,$ the equation with our
specific choice of $D $  can be  converted to
\begin{equation}  
 6 \exp{(\sigma X/2)} H_X + H H_{\xi} + H_{\xi \xi \xi} = (\frac{21 \sigma}{2}) \exp{(\sigma X/2)}
H.
\label{special1}
\end{equation}
Defining a new coordinate variable  as 
$T = -\frac{\exp{(-\sigma X/2)}}{3 \sigma}$, equation (\ref{special1}) 
can be transformed now to a  convenient form of the
so called  concentric KdV equation 
\begin{equation}
 H_T + H H_{\xi} + H_{\xi \xi \xi} + \frac{7}{2T} H = 0,
\label{cKdV1}
\end{equation}
which is a known integrable equation derivable from the hydrodynamic
equations with cylindrical symmetry \cite{Johnson}.
An  exact solution of the  variable
coefficient  KdV equation (\ref{cKdV1}) is presented in \cite{Johnpillai} in  the
rational form as \ \ 
$H = \frac{(c - \frac{5}{2} \xi)}{T}.$
 Using the relation with our original
field: $\eta_0 = \frac{D^2}{9} H$ and reverting
to our old coordinates $\xi, X$ we can transform back the solution 
to obtain the required exact solution for the surface wave 
\begin{equation}
 \eta_{0} = -\frac{\sigma}{3}(c - \frac{5 \xi}{2}) \exp{(-3 \sigma X/2)},
\label{RationalF}
\end{equation}
with an  arbitrary constant $c$. 
Note that this is a rational solution, not of solitonic type and it
 behaves differently for  different values of $\xi .$
For $\xi < \frac {2c} {5}, \ \  $  $\eta_0 <0 $, for  $\xi > \frac {2c} {5}, \ \  $  $\eta_0
>0 $, while at  $\xi = \frac {2c} {5}, \ \  $  $\eta_0
=0 $ (see Figure 6 ).   
Solution (\ref{RationalF}) shows that, the amplitude decays down
due to the exponential damping factor,  as the wave propagates along the
positive  $X$
direction.
Thus the surging waves are controlled to damping wave through balancing with
the leakage at the bottom as we have aimed at.
 At $ \xi \to \pm \infty$ the wave profile shows
 divergent nature. However since our intention is to consider the wave
propagation towards the shore the damping effect obtained along $X$
is the relevant factor.   
\begin{figure}[!h]
\centering

{
 \includegraphics[width=7 cm, angle=0]{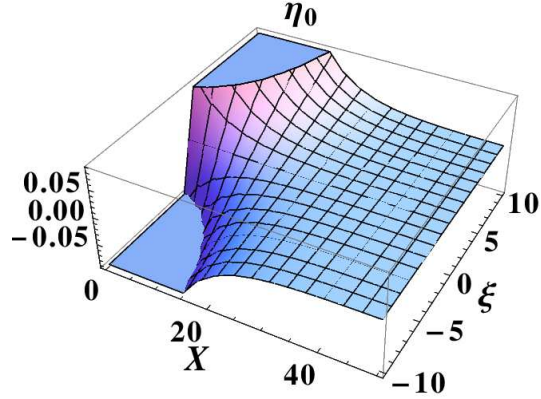}
}

\caption{3D  plot of the exact wave solution (\ref{RationalF}) in the
$\xi, X$
plane with exponentially decreasing depth with $X$ and a bottom leakage with $\sigma
=0.1$.
The  amplitude decay with the distance traveled along $X$, is evident. The divergent
nature of the solution in  $\xi $ can be detected from the figure.
}
\label{fig:6}
\end{figure}
\subsection{Time dependent leakage:}
In all the previously discussed cases, the leakage function $g$ is assumed to depend slowly on the space
variable $x$ as $g(\epsilon x)$. As an extension to the problem,
 we consider here a special kind of leakage function, which is depends slowly on time, as $g(\epsilon t)$.
When the bottom boundary depends on time, then the two dimensional potential 
flows of an ideal fluid with a free surface are considered 
 in \cite{PLAtd} .

Here, the bottom boundary condition at the variable bathymetry $z = B(X)$ becomes
 \begin{equation}
 w = u \frac{d b}{ d x} - \epsilon  g(\epsilon t) G(\eta).
\label{tdBBC}
\end{equation}
 where the leakage function $g$ depends slowly on time $t$.
As we have mentioned we assume $G = \eta$ to get the maximum benefit
of damping due to leakage.
 For detailed investigation we introduce a new set of variables
\begin{eqnarray}
 \xi = \frac{1}{\epsilon} \chi(X)  - t,   \ \
X = \epsilon x,
\Theta = \epsilon t
\label{tdstretched}\end{eqnarray}

Note that, here we have introduced a new slow time variable $\Theta$ which also depends slowly on time.
After a bit of calculations 
  we can finally derive the 
 surface wave evolution equation  
\begin{equation}
 2 \sqrt{D} \eta_{0X} + \frac{3}{D}\eta_0 \eta_{0 \xi} + 2 \eta_{0\Theta} + (\frac{D'}{2
\sqrt{D}}+ g(\Theta))\eta_0
+ D \eta_{0 \xi \xi \xi} = 0.
\label{vckdvtd}
\end{equation}
Note that two extra terms arise due to the slow time $\Theta$ which can be canceled 
in the following way.
  
Let us consider a new transformation $\eta_0 = f(\Theta) \phi(\xi, X)$.
We consider $g(\Theta)$ to be such that the extra two terms which arose due to the slow time $\Theta$
cancels each other such that
\begin{equation}  
 2 \eta_{0\Theta} + g(\Theta)\eta_0 = 0
\end{equation}
which finally gives $f = A \exp^{-\frac{1}{2}\int g d\Theta}$, where $A$ is a constant.
The equation satisfied by the function $\phi$ is nothing but that obtained by Johnson (\ref{kdvJohnson}).
Hence using their solution (\ref{surging}) as given in \cite{Johnson} the final solution can be written as

\begin{eqnarray}
 \eta_0 = \frac{A}{D}
\exp^{-\frac{1}{2}\int g d\Theta}
sech^2 {[\sqrt{\frac{3 A_0}{4 D^3}}(\xi -
\frac{D^{-(\frac{5}{2})} A_0 X}{2})]},
\label{td}
\end{eqnarray}

The dynamics of the solution (\ref{td}) can be explained like follows. As the wave propagates towards
the shallower region, due to the factor $1/D$ the wave amplitude increases, whereas due to the exponentially
decaying factor, which depends on time the amplitude increase is compensated to some extent.
But the leakage function $g(\epsilon t)$ should be synchronized in such a way that as the wave starts increasing
it starts working. Such physical mechanism and installations can be used in the other physical situations as
required.

\end{document}